\documentclass{article}
\usepackage{cite} 
\usepackage[square,sort,comma,numbers]{natbib}
\usepackage{spconf,amsmath,graphicx}

\usepackage{balance} 
\usepackage{lipsum}
\usepackage{subcaption}
\usepackage{graphicx}
\usepackage{amsmath}
\usepackage{makecell}
\usepackage{rotating}
\usepackage{colortbl}
\usepackage{multirow}
\usepackage{adjustbox}
\usepackage{amsfonts}
\usepackage{url}
\usepackage[ruled,vlined,linesnumbered]{algorithm2e}

\title{Optimal Multi-Agent Path Finding 
for\\Precedence Constrained Planning Tasks}

\name{Kushal Kedia*, Rajat Kumar Jenamani*, Aritra Hazra, Partha Pratim Chakrabarti}
\address{Indian Institute of Technology Kharagpur}

\newcommand{\BibTeX}{\rm B\kern-.05em{\sc i\kern-.025em b}\kern-.08em\TeX}

\begin{document}


\newcommand{\kushal}[1]{ {\color{red}#1} }
\newcommand{\rajat}[1]{ {\color{blue}#1} }
\newcommand{\aritra}[1]{ {\color{green}#1} }
\newcommand{\ppc}[1]{ {\color{purple}#1} }
\newcommand{\rulesep}{\unskip\ \vrule\ }



\maketitle 

\begin{abstract}
Multi-Agent Path Finding (MAPF) is the problem of finding collision-free paths for multiple agents from their start locations to end locations. We consider an extension to this problem, Precedence Constrained Multi-Agent Path Finding (PC-MAPF), wherein agents are assigned a sequence of planning tasks that contain precedence constraints between them. PC-MAPF has various applications, for example in multi-agent pickup and delivery problems where some objects might require multiple agents to collaboratively pickup and move them in unison. Precedence constraints also arise in warehouse assembly problems where before a manufacturing task can begin, its input resources must be manufactured and delivered. We propose a novel algorithm, Precedence Constrained Conflict Based Search (PC-CBS), which finds makespan-optimal solutions for this class of problems. PC-CBS utilizes a Precedence-Constrained Task-Graph to define valid intervals for each planning task and updates them when \textit{precedence conflicts} are encountered. We benchmark the performance of this algorithm over various warehouse assembly, and multi-agent pickup and delivery tasks, and use it to evaluate the sub-optimality of a recently proposed efficient baseline.
\end{abstract}


\renewcommand*{\thefootnote}{\fnsymbol{footnote}}
\footnote[1]{The first two authors contributed equally to this work.}
\setlength{\AlCapSkip}{1em} 


\section{Introduction} \label{sec:introduction}


The classical Multi-Agent Path Finding (MAPF) problem considers the case where each agent has to solve one planning task while avoiding collisions amongst themselves. Many real world applications of MAPF require agents to complete more than one task. Recent work has addressed Multi-Agent Pickup and Delivery (MAPD) \cite{mapd} which aims to solve sequential planning tasks for a group of agents. The sequential nature of tasks in MAPD gives rise to in-schedule dependencies for each agent. Further, cross-schedule dependencies between agents can develop due to precedence constraints which require completion of certain tasks by other agents as a prerequisite for starting other dependent tasks. The class of problems with inter-task precedence constraints, defined as Precedence-Constrained Multi-Agent Path Finding (PC-MAPF), models a variety of scenarios in warehouse applications. In this paper, we are motivated by the use cases of PC-MAPF in warehouse assembly and collaborative MAPD.  



In the former setting, one can consider a warehouse with manufacturing stations. Agents are required to transfer materials between manufacturing locations while avoiding collisions. The overall assembly problem comprises sub-assembly components which depend on each other. Temporal precedence constraints exist between planning tasks because each  manufacturing station can generate its output only after receiving its required input materials. 
 
The other setting is a generalised formulation of MAPD where objects require more than one agent for their transport. Here, agents must arrive at the object pickup location before moving in unison towards the drop-off point. Thus, the beginning of the transport is constrained by the arrival of each collaborating agent at the pickup location. Recent work \cite{drone_1, drone_2} has demonstrated the feasibility of low-level control required for collaborative MAPD, using multiple drones to lift objects that cannot be picked up by a single drone.
 
 \begin{figure}[t!]
    \centering
    \includegraphics[width=0.99\linewidth]{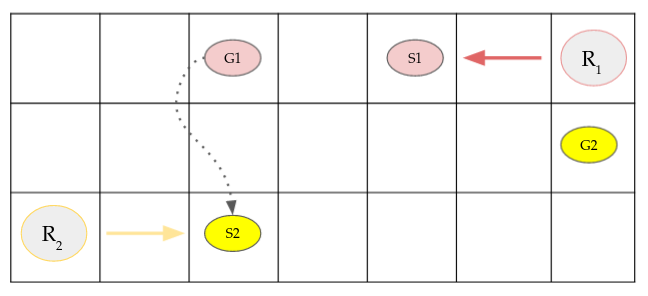}
    \vspace{-0.25cm}
    \caption{{\small Example PC-MAPF problem where agent $R_1$ is assigned a transport task from $S1$ to $G1$ and agent $R_2$ is assigned a transport task from $S2$ to $G2$. $R_2$ can pickup the object at $S2$ only after $R_1$ completes delivery at $G_1$. (represented by a precedence constraint edge)}}
    \label{fig:introduction}
\end{figure}

Conflict-Based Search (CBS) \cite{cbs}, a state-of-the-art optimal algorithm for MAPF, finds shortest paths for each agent by updating their solution paths iteratively to avoid collision conflicts between agents. Due to the additional cross-schedule dependencies, an optimal algorithm for the PC-MAPF problem domain will also have to consider the valid intervals for starting each planning task. The algorithm proposed by \cite{hcbs} is closest to our work and attempts to solve the PC-MAPF problem optimally using a hierarchical algorithm based upon CBS. We shall refer to this algorithm as Hierarchical-CBS (H-CBS). H-CBS builds plans for a group of agents by finding collision-free paths for each task segment individually which are consistent with inter-task precedence constraints. However, as noted by \cite{brown2021algorithms}, due to segmented nature of planning, H-CBS is sub-optimal for problems requiring \textit{pre-emptive delays} (example shown in Fig. \ref{fig:sub-opt}). In such cases, though there is no constraint on the plan of an individual task, delaying an agent's plan pre-emptively can shorten the duration of the overall plan.

 \begin{figure}[t!]
    \centering
    \includegraphics[width=\linewidth]{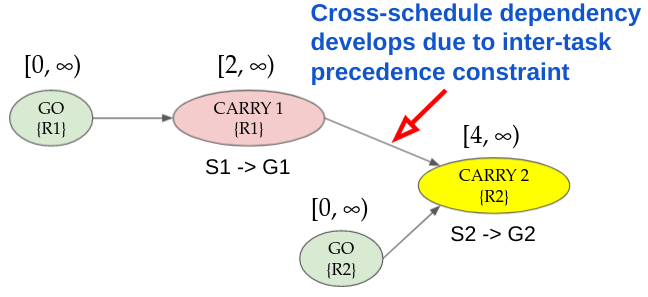}
    \vspace{-0.7cm}
    \caption{{\small Task Graph corresponding to Fig \ref{fig:introduction}. GO nodes represent movement to pickup locations \& CARRY nodes depict the transport tasks. Agents assigned to each task are indicated. The possible valid interval for starting each task is displayed (without considering inter-agent collisions). Despite $R_2$ being able to start CARRY-2 at timestep 2, CARRY-1 cannot finish before timestep 4. Hence, the valid interval for starting CARRY-2 begins from timestep 4.}
    \label{fig:task_graph}}
\end{figure}
As shown in Figure \ref{fig:task_graph}, any PC-MAPF can be represented a Task Graph, where each node denotes a planning task to be completed by a set of agents from a start location to an end location. The directed edges denote precedence constraints between tasks, i.e, all preceding tasks must have been completing before a task is allowed to begin. In this paper, we present the first algorithm to solve any PC-MAPF problem  optimally, \textbf{Precedence  Constrained  Conflict Based  Search (PC-CBS)}, a two-level algorithm comprising a Conflict Tree (C.T) at its high level and a multi-waypoint A* at the low level. At the high level of the algorithm is defined the task graph for the problem along with valid intervals of each task. As in CBS, the plans of agents are constrained and updated in a tree-like fashion when conflicts are encountered. Apart from collision conflicts which constraint the plans of each agent, we additionally introduce \textit{precedence conflicts}, which arise when agents' plans are inconsistent with the defined task graph for the problem. These conflicts constraint the valid intervals for each task in the Task Graph. The low-level search of the algorithm employs an optimal multi-waypoint A* search which finds paths for each agent separately considering the start and end locations of each planning task as waypoints. Like in H-CBS, the low-level search makes use of tie-breaking heuristics \cite{tieastar} to choose paths which are likely to reduce conflicts in future iterations of CBS. Our key contributions are summarized below:



\begin{itemize}
    \item We show that any PC-MAPF problem can be represented as a Task Graph with valid intervals for starting each task.
    \item  We propose, \textbf{PC-CBS}, the first optimal algorithm to find minimum-makespan solutions for any PC-MAPF problem.
    \item We benchmark the performance of PC-CBS in warehouse assembly and collaborative MAPD problem instances and evaluate the sub-optimality of a recently proposed baseline.
\end{itemize}
The rest of the paper is organized as follows. Section~\ref{sec:related_work} presents existing works in this direction. We formulate the PC-MAPF problem setup and its task graph representation in Section~\ref{sec:formulation}. In Section~\ref{sec:approach}, we detail our proposed methodology leveraging search algorithms and prove its optimatility. Further, we discuss the Hierarchical-CBS (H-CBS) algorithm in Section~\ref{sec:hcbs}. Section~\ref{sec:experiment} presents the experimental results demonstrating the performance of our proposed method. Finally, Section~\ref{sec:conclusion} concludes the paper.

\section{Related Work} \label{sec:related_work}

Despite the MAPF problem with fixed goals proven to be NP-Hard \cite{nphard} under the sum of costs and makespan objectives, there have been effective attempts \cite{sharon_increasing_2013, mstar} at building optimal algorithms for this problem. Among them, a tree-based algorithm, Conflict Based Search (CBS) \cite{cbs}, has seen wide applications in MAPF. In recent literature, various improvements \cite{icbs} have been proposed on top of CBS including extensions to continuous time domains \cite{ccbs}, and bounded-subpotimal algorithms \cite{subopt_cbs}. Since MAPF is a special case of the PC-MAPF problem without inter-task precedence constraints, we claim that solving PC-MAPF optimally for the same objectives are also at least NP-Hard. Our work builds upon the CBS algorithm. 

Instead of assuming fixed goals, the Anonymous MAPF (AMAPF) is the problem of assigning goals for each agent. \cite{amapf} show that AMAPF is no longer NP-Hard under the makespan objective and can be reduced to a maximum-flow problem. TA-CBS \cite{tacbs} uses a high-level search forest instead of a tree to optimally assign goals to agents while minimizing the sum of costs over all the tasks. They also propose a bounded sub-optimal version of their algorithm, which scales gracefully with the size of the problem. The popular M* \cite{mstar}, an optimal MAPF algorithm was extended in \cite{mstarext} by integrating it with task reassignments. Co-CBS \cite{co-cbs} integrates a cooperation-planning module into CBS and introduces the problem of active collaboration by assigning "meeting" points for transfer of goods in a warehouse setting.

There have been numerous works in finding collision-free paths for a sequence of pickup and delivery tasks. TCBS \cite{tcbs} solves the combined task allocation and path-finding problem where agents can be assigned a sequence of tasks. However, the performance of the algorithm significantly degrades upon an increase in the number of tasks. Two offline algorithms for multi-agent pickup and delivery have been proposed by \cite{ta_mapd}, both of which first assign a sequence of tasks to agents by solving a special Travelling Salesman Problem (TSP) that does not consider collisions between agents. These algorithms then plan for collision-free trajectories for agents that solve the assigned task sequences. \cite{life_mapd} consider an online setting where new pickup and delivery tasks can arrive. It proposes two decentralized algorithms to solve this problem that performs prioritized search over agents using token passing. However, these algorithms are sub-optimal. Other work \cite{mlastar} attempts to solve this problem by using a Multi-Level A* supported by a novel heuristic to find the shortest path for agents given an ordered list of goals. 

We are interested in solving MAPF with precedence constraints between planning tasks. In the taxonomy for multi-robot task allocation listed by \cite{taxo}, PC-MAPF belongs to a wide class of problems having cross-schedule dependencies. We have considered the precedence constrained case where completion of a task is pre-requisite for another, but cross-schedule dependencies can also arise when tasks need to be synchronized or require time-windowing, i.e, a task must be completed while a larger task is in progress. \cite{cross_sched_1, cross_sched_2} analyze synchronization and precedence constraints commonly arising in vehicle routing and scheduling problems. However, an assumption made in this line of work is that collisions do not arise between agents, which is in contrast to the MAPF literature.

A recurring theme in Task Allocation and MAPF (TA-MAPF) consists of solving a relaxed task assignment problem, for example without collisions, and then finding feasible solution paths for agents given a particular assignment. The work by \cite{hcbs} too, which is closest to ours, at the top level solves a task assignment problem without considering collisions between agents. For finding solutions to the PC-MAPF problem formed after task assignment, it proposes H-CBS. However, this algorithm has been shown to be sub-optimal for a class of pre-emptive delay problems by \cite{brown2021algorithms}. To the best of our knowledge, ours is the first work which optimally solves the PC-MAPF problem. Since the focus of this paper is to find optimal solutions given the task assignment, we use a greedy strategy to assign tasks to agents during problem generation. Our PC-CBS algorithm can be easily extended by a task assignment module.

\section{Problem Formulation and Notation} \label{sec:formulation}

We first discuss the classical MAPF problem, and then describe the specifications added to it in the general PC-MAPF problem. 

\subsection{Classical MAPF}
In classical MAPF, the configuration space each agent is represented by an undirected graph $G=\{V, E\}$, where $V$ is the set of vertices in $G$, each of which represents a valid  configuration for the agent, and $E$ is the set of edges, each representing possible movements from one configuration to another for each agent. Time is assumed to be discretized into time steps, and the an agent can move over any edge in one time step. The team of agents, $A = \{a_1, \dots , a_N\}$, is assigned a start configuration, $S = \{s_1, \dots , s_N\} \in V$ and goal configuration, $G = \{g_1, \dots , g_N\} \in V$. All agents start at timestep $0$ from their start configurations. A feasible joint path for the problem is defined by $\pi = \{\pi_1, \dots , \pi_N\} $, where an individual path $\pi_i = \{v^1_i, \dots, v^k_i\}$ describes a sequence of configurations for agent $a_i$ from $s_i$ to $g_i$ which avoids collisions with the paths of other agents.  The path cost of an individual agent is the time taken to reach the goal. An optimal solution optimizes an objective function over the joint path. Common objective functions include the sum of costs, which adds up the path costs for each agent, and the makespan, which considers the maximum path cost among all agents. Two agents are said to be in a collision if they occupy the same vertex in $V$ at the same timestep, or if they exchange positions between timesteps, i.e, two agents move from opposite ends of the same edge represented by $v_j \leftrightarrow v_j, v_{i,j} \in V$, they are said to be in collision.

\subsection{PC-MAPF}
We now define PC-MAPF which incorporates multiple planning tasks for each agent. As in classical MAPF, paths are planned over an undirected graph, $G=\{V, E\}$ with unit edge costs, and agents are defined an initial starting configuration $S$. We further define goal configurations for each agent that act as final \textit{parking} locations, i.e, their paths must end at these locations. Additionally, a set of pickup and delivery tasks $T = \{t_{1} \dots t_{m}\}$ are required to be completed by a team of agents. The pickup and delivery locations of tasks are called task
endpoints. Each agent $a_{i}$ is assigned a sequence of tasks that it must complete, called a task allotment list. There exists a set of precedence constraints between tasks, which is defined as $t_i \rightarrow t_j$ denoting that $t_i$ must be completed before $t_j$ can begin. Each task $t_{i}$ $\in$ $T$ is a tuple $(R_{i}, s_{i}, g_{i})$ where $R_{i}$ is the set of agents allocated to this task, $s_{i}$ denotes the pickup location of the task, and $g_{i}$ denotes the delivery location. Even when there are no explicit constraints between tasks, precedence constraints exist during the planning of a task requiring more than one agent for pickup. All agents collaborating for the task have to arrive at the pickup location before delivery can be initiated.

A feasible solution path in PC-MAPF must avoid collision and precedence conflicts between agents. An important distinction to classical MAPF is that here agents are said to be in a collision if they occupy the same vertex in $G$ and are not carrying out the same pickup and delivery task, i.e, the task required more than one agent. To avoid precedence conflicts, for each task $t_i \in T$, the set of agents allocated to this task, $R_{i}$ must start pickup at the same timestep and tasks which are precedent to $t_i$ must have been completed before $t_i$ begins. In this work, we find makespan optimal solutions to PC-MAPF problems which is equivalent to the last timestep at which an agent arrives at their final parking location.

\begin{figure}[t!]
    \centering
    \includegraphics[width=0.7\linewidth]{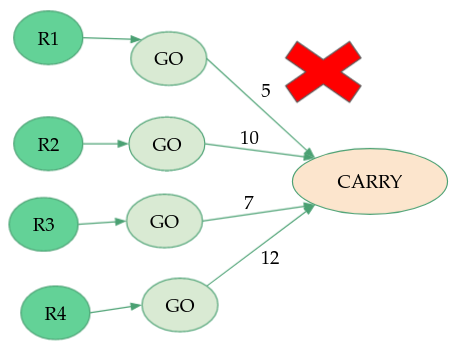}
    \vspace{-0.25cm}
    \caption{\small{Agents $R\{1-4\}$ begin at their distinct initial locations at timestep $0$ and start a collaborative "carry" task at different timesteps resulting in a precedence conflict}}
    \label{fig:pred_conflict}
    \vspace{-0.5cm}
\end{figure}
\subsection{Task Graph Representation}

A directed acyclic graph (DAG) can be constructed to define any PC-MAPF problem called the Task Graph, $G_T = \{V_T, E_T\}$. Each vertex of the graph $v \in V_T$ represents any planning task consisting of a set of agent from one position to another. This may either be a "CARRY" vertex which is a pickup and delivery task called a or a "GO" vertex, requiring the agent to arrive at the pickup location of a transport task. A directed edge, $(v \rightarrow u) \in E_T$ represents a precedence constraint which means that $v$ must be completed before $u$ can start. Each agent is assigned a task allotment list, which contains a sequence of pickup and delivery tasks it must execute. Along with the "GO" tasks the agent must execute to reach the pickup points, these tasks are all connected by edges of the Task Graph and act as precedence constraints. 

In Collaborative MAPD, multiple "GO" tasks arrive at the same pickup and delivery task vertex (eg. Fig \ref{fig:pred_conflict}). The end point of the "GO" task is same as the start point of the successor task. However, precedence constraint edges are not required to be connected with their task end points. These edges may also represent the arrival of input resources around manufacturing stations after which generation of output goods can begin. For each planning task vertex, we further define two intervals at its endpoints within which an agent has to start and end that particular task. If the start point of a task has an interval $(t_{min}, t_{max})$, agents while executing this task can occupy positions at the start only within this time interval. An example Task Graph has been shown in Fig \ref{fig:task_graph}.

\section{Proposed Approach} \label{sec:approach}

Conflict Based Search (CBS) \cite{cbs} is a complete and optimal solver for classical MAPF. It solves a given problem by finding plans for each agent separately, detecting collision conflicts between these plans. The conflicts are resolved by replanning for the individual agents subject to specific constraints. CBS runs two search algorithms --- a low-level search algorithm that finds paths for individual agents subject to a given set of constraints and a high-level search algorithm that maintains a binary tree called the conflict tree (CT). Each node of the CT chooses conflicts to resolve, and divides the node into two sub-nodes. If there are no more conflicts to resolve in a node, it is said to be a leaf node which contains a feasible solution path to the problem. The CT is searched in a best-first manner based on the makespan cost of a node.

PC-CBS extends CBS to PC-MAPF problems. We note that CBS assigns collision constraints only to specific agents.
Our key contribution is the introduction of precedence conflicts between tasks and constraints that arise from these conflicts. While collision conflicts arise when the position of agents at the same location are intersecting in time, precedence conflicts arise where the combined plan of the agents does not adhere to the Task Graph of the problem. The constraints are imposed at the task level and apply to all agents. The low-level search of PC-CBS ensures that agents adhere not only to the collision constraints imposed on them, but also begin and end tasks within defined intervals.

An example of a precedence conflict is shown in Fig \ref{fig:pred_conflict}. Here, collaborating agents pick up an assigned object at different timesteps in their individual plans. In order to resolve precedence conflicts between agents, PC-CBS branches out into nodes that contain constraints over task intervals that aim to invalidate the earlier found joint plan. These constraints update defined valid intervals of the task's end-points, and also influences the intervals of tasks connected to this task node in the Task Graph. In this section, we first explain the high level search for building the conflict tree and then we show how the constraints are imposed on the low level multi-waypoint A* search by precedence conflicts. Finally, we prove that our proposed algorithm is complete and optimal.



\subsection{High Level Search}

\newcommand\mycommfont[1]{\footnotesize\ttfamily\textcolor{blue}{#1}}
\SetCommentSty{mycommfont}

\begin{algorithm}[t]
  \caption{ High - Level Search of PC-CBS}
  \SetKwInOut{Input}{Input}

  \Input{ Task Graph $G_T = \{V_T, E_T\}$ }
  
  $Root$.constraints $\leftarrow$ $\emptyset$ \; 
  $Root$.intervals $\leftarrow$ Initialize Valid Task Intervals()  \;
  $Root$.solution $\leftarrow$ invoke low-Level search for each agent $r_{i}$ \tcp*{find individual plans for agent using low level search in decreasing order of estimated execution times}
  insert $Root$ in OPEN list \;
    \While{solution not found}{
        P $\leftarrow$ pop best node from OPEN \tcp*{lowest makespan cost}
        \For{$t_i$ $\in$  topological\_sort($G_T$)}{
            \uIf{$t_i\ consistent\ with\ G_T$}{ 
                continue \;
            }
            children $=$ Resolve\ Precedence\ Conflict() \;
            OPEN.insert(children) \; 
            Goto next iteration of while \;
        }
        Validate paths in P until a collision conflict\;
        \uIf{P has collision conflict}{
            ($R_{p}$, $R_{q}$, $v$, timestep) $\leftarrow$ first collision conflict \;
            children $=$ Resolve Collision Conflict() \;
            OPEN.insert(children) \; 
            continue \; 
        }
        return P.solution \tcp*{P is goal node} 
    }
\end{algorithm}

The High Level Search is a Conflict Tree (C.T). A node $N$ of the PC-CBS C.T is defined by three attributes:
\begin{enumerate}
  \item $N.constraints \rightarrow$ Set of vertices or edges invalidated for agents at particular timesteps.
  \item $N.intervals \rightarrow$ Valid intervals for endpoints in each planning task of the Task Graph.
  \item $N.solution \rightarrow$ Solution paths for each agent which are consistent with the constraints and task intervals.
\end{enumerate}
At the root node (Lines 1-3 in Alg. 1), there are no constraints and the valid intervals for each task are initialised by solving the planning problem without considering collisions with each other. Hence, all task endpoints have a defined minimum time before which they cannot be reached but there is no maximum bound yet, i.e, all intervals have a lower bound and an infinite upper bound. The solution paths are then computed by invoking the low level search. The tree is searched using a best-first strategy, expanding the nodes whose solution paths have the lowest makespan, thus ensuring makespan-optimal solutions.

Following the work on prioritized motion planning by \cite{prioritizedMP}, whenever paths have to be computed for multiple agents, the low-level search is called in decreasing order of the estimated execution times of their task sequences. For the first calls of the low-level search, the estimate of an agent's total execution time is the minimum time at which an agent can complete the last task in its itinerary. In following iterations, whenever a child node in the C.T is expanded, the path costs in the parent's solution paths are used for this purpose. The tie breaking heuristics of the low-level search find solutions with the minimum number of conflicts provided the path does not increase the makespan cost of the overall plan. This allows agents whose path costs are low to delay their plans in order to avoid conflicts with other agents.

Conflicts are detected in a hierarchical manner by the high-level search which first checks the path for precedence conflicts. If there are any present, it splits into child CBS nodes that inherit the constraints of their parents along with additional constraints that aim to invalidate the plan found by its parent. The node is first checked for the existence of any precedence conflicts. If there exists a predecessor for any task which ends after it begins, a precedence conflict is detected and the intervals for this task are updated by Algorithm 2. Once a node does not have any more precedence conflicts, it is checked for collision conflicts. In case collision conflicts are present, it splits with nodes with collision constraints on the agents. Prioritization of precedence conflicts over collision conflicts is essential to the validity of the algorithm as it ensures that the paths of individual agents carrying out the same collaborative task do not differ and the Task Graph is not violated.

\vspace{-0.3cm}
\subsection{Resolving Precedence Conflicts}
Similar to how CBS resolves collision conflicts, we aim to eliminate invalid solutions from the search space. This is done by changing the valid intervals for starting or ending any task. For example, consider the precedence conflict in Fig \ref{fig:pred_conflict} where the agents be assigned the same collaborative carry task. Due to being at different initial locations, the first node expanded in by the high-level search returns a set of paths where the agents execute pickup at different timesteps - namely $R1$ at $t=5$, $R2$ at $t=10$, $R3$ at $t=7$, and $R4$ at $t=12$. As there is a precedence conflict, this CBS node is invalidated and its child nodes must be generated. We propose to split by generating two child nodes which breaks the start point intervals of this task into two disjoint intervals (Algorithm 2). Any timestep can be chosen for splitting this interval as long as the original interval shrinks. Empirically we find that taking the maximum timestep among the predecessor tasks works the best. The new C.T. node whose interval is upper bounded by the maximum time of the preceding task is often infeasible and gets killed in the next iteration of invoking the low-level search. This keeps the branching factor of the tree in check while mantaining completeness. In \ref{fig:pred_conflict}, let's say the initial valid interval for starting the CARRY task is $[0, \infty)$. This interval is broken down into $[0, 11]$ and $[12, \infty)$ by our expansion procedure.

\begin{algorithm}[t]
    \caption{Resolve Precedence Conflict}
    \SetKwInOut{Input}{Input}
    \Input{ Task Graph $G_T$, Parent Node P, Conflict Task T}
    split\_timestep = max end-time for Pred(T) \tcp*{for predecessors of T in $G_T$, find the maximum timestep violation}
    old\_interval = P.intervals[T].start \tcp*{extract previous valid interval of start point of task from parent node}
    \uIf{split\_timestep $>$ old\_interval.maxTime}{
        split\_timestep = old\_interval.maxTime\;
    }
    maxInterval = [old\_interval.minTime, split\_time-1] \;
    minInterval = [split\_time, old\_interval.maxTime] \;
    Spawn children nodes with updated Intervals \;
    return children \;
\end{algorithm}

It is noted that dividing the valid intervals of a task into disjoint sets preserves all possible sets of solutions. However, the effectiveness of CBS is achieved by being able to delete invalid solutions from the search space. PC-CBS does this while resolving precedence conflicts by exploiting the structure of the Task Graph. By updating the task interval of one task, it is possible to deduce update rules for the task intervals of other tasks as shown in Algorithm 4. In Lines 1-6, the task graph is iterated in topological order. For each task node it is inferred that its minimum starting time must be greater than or equal the minimum end-time of all its predecessors. Further, we know the minimum cost required to complete each task is the path length of a task without considering any collisions. This is used to update the end-time. Similarly, in Lines 7-12, the maximum times of each task are updated via a reverse topological sort by leveraging the fact that the maximum end-time of a task must be less than or equal to the maximum start-time of a successor task.

\begin{algorithm}[t]
  \caption{ Update Task Intervals}
  \SetKwInOut{Input}{Input}
  \SetKwInOut{Output}{Output}
  \Input{C.T. Node $N$}
  
    \For{$t_i$ $\in$  topological\_sort($G_T$)}{ 
        \For{$t_{pred}$ $\in$  predecessors($t_i$)}{
            \uIf{N.intervals[$t_{pred}$].endpoint.minTime > N.intervals[$t_i$].startpoint.minTime }{
                N.intervals[$t_i$].startpoint.minTime = N.intervals[$t_{pred}$].endpoint.minTime \;
            }
        }
        \uIf{N.intervals[$t_i$].startpoint.minTime + min-cost($t_i$) > N.intervals[$t_i$].endpoint.minTime}{
            N.intervals[$t_i$].endpoint.minTime = N.intervals[$t_i$].startpoint.minTime + min-cost($t_i$) \;
        }
    }
    \For{$t_i$ $\in$  reverse\_topological\_sort($G_T$)}{
        \For{$t_{succ}$ $\in$ successors($t_i$)}{
            \uIf{N.intervals[$t_{succ}$].startpoint.maxTime < N.intervals[$t_i$].endpoint.maxTime }{
                N.intervals[$t_i$].endpoint.maxTime = N.intervals[$t_{succ}$].startpoint.maxTime \;
            }
        }
        \uIf{N.intervals[$t_i$].endpoint.maxTime - min-cost($t_i$) < N.intervals[$t_i$].startpoint.maxTime}{
            N.intervals[$t_i$].startpoint.maxTime = N.intervals[$t_i$].endpoint.maxTime - min-cost($t_i$) \;
        }
    }
    return N \;
\end{algorithm}

\subsection{Low Level Search}
The primary goal of the low-level search is to find a path for an agent that adheres to its collison conflicts and respects the valid intervals for each of its assigned tasks. If an agent executes the sequence of actions in its path starting from its initial location, then it will pick up and deliver each of the object tasks assigned to it in a sequential manner. This can be done by running a variation of A$^*$ planning algorithm which is forced to visit a sequence of waypoint configurations in its path. Let the set of ordered tasks of size $K$ assigned to agent a be $\{t_{1}^{a}, t_{2}^{a}, … , t_{K}^{a}\}$. The state representation for path finding for is \{ position, timestep, tasks\_completed\}, where:

\begin{enumerate}
    \item position $\rightarrow$ the configuration of agent A in the graph
    \item timestep $\rightarrow$ the current timestep of A
    \item tasks\_completed $\rightarrow$ the set of completed tasks $\{t_{1}^{a}, t_{2}^{a},\dots, t_{k}^{a}\}$ 
\end{enumerate}

The low level search problem for agent $a$ which starts at $v_{s}^{a}$ and has been assigned the task sequence $\{t_{1}^{a}, t_{2}^{a}, … , t_{K}^{a}\}$ then becomes to find the minimum cost path from start to goal where, start = ($v_{s}^{a}$,  0, $\emptyset$) and, the goal condition for a node \{ position, timestep, tasks\_completed = $\{t_{1}^{a}, t_{2}^{a},\dots, t_{k}^{a}\}$\} is that $k = K$ (all assigned tasks have been completed). However, the ultimate objective of PC-CBS is to find makespan-optimal paths for all agents. We use this insight to make use of possible delays in paths of an agent to avoid conflicts which would have been detected in future iterations of the high level search. A tuple of heuristics are defined for this purpose in Algorithm 4. These are checked from the first to last, with the later values checked only if there is a tie. This cascaded tie-breaking procedure \cite{tieastar} has been successfully used to tune the efficiency of A* search. In our case, however, we are interested in using the heuristics to guide search towards solutions which avoid conflicts. 

The first heuristic, called the delay cost, is the minimum time that the path is guaranteed to delay the entire plan. This heuristic maintains the makespan-optimality of the algorithm. The second and third heuristics aim to reduce the number of precedence and collision conflicts by preemptively comparing paths with that of other agents. The fourth heuristic is the variation of the standard $f-value$ used in $A*$. It uses $moveActions$ as the cost-to-come measure which is equivalent to the distance travelled by the agent instead of travel time. Finally, paths with lower cost-to-go measures are prioritised over nodes with lower $f-values$. The cost-to-go function is looked up from a pre-computed table of shortest path costs between any two vertices. This is obtained as a pre-processing step by running the Floyd-Warshall algorithm \cite{floyd} on the configuration space graph without considering collisions with other agents.

\begin{algorithm}[t]
  \caption{Heuristics for Low Level Search }
  \SetKwInOut{Input}{Input}
  \SetKwInOut{Output}{Output}

  \Input{Plans P, Current node v, Current path p}
  
	$C1 = max(CostToCome(v) + CostToGo(v) - makespan , 0)$\;
    $C2 = countPrecedenceConflicts (p, P)$\;
    $C3 = countCollisionConflicts (p, P)$\;
    $C4 = MoveActions (p) + CostToGo(v)$\;
    $C5 = CostToGo(v)$\;
    $C6 = CostToCome(v) + CostToGo(v)$\;

    return $(C1, C2, C3, C4, C5, C6)$;
    
\end{algorithm}

\subsection{Proof of Optimality}

In this section, we prove the optimality of PC-CBS.

\noindent \textbf{[Definition 1]} \textit{For given node $N$ in the C.T., let $CV (N)$ be the set of all solutions that are consistent with the collision constraints and task intervals of $N$ and are don't have any collision or precedence conflicts.}

If $N$ is not a goal node, then the solution at N will not be part of $CV (N)$ because it is not valid, i.e, it has a conflict. 

\noindent \textbf{[Definition 2]} \textit{For any solution $p \in CV (N)$ we say that node $N$ permits the solution $p$ if it is consistent with its constraints.}

The root node having no constraints permits all valid solutions. The cost of a solution in $CV (N)$ is the lowest makespan cost among the set of its permitted paths. Let $minCost(CV (N))$ be the minimum cost over all \textsl{valid} solutions in $CV (N)$.

\noindent \textbf{[Lemma 1]} \textit{The cost of a node $N$ in the CT is a lower bound on $minCost(CV (N))$.}

\noindent \textit{Proof:} $N.cost$ is the lowest makespan cost among the set of paths that satisfy the collision constraints and task intervals of $N$. These set of paths need not be valid. Thus, $N.cost$ is a lower bound on the cost of any set of paths that make a valid solution for $N$. \hfill $\Box$

\noindent \textbf{[Lemma 2]} \textit{ For a valid solution p, there exists a CT node N in OPEN List that permits p at all time.}

\noindent \textit{Proof:} For the base case, OPEN only contains the root node, which has no constraints. Consequently, the root node permits all valid solutions and also $p$. The statement holds true iff when branching at a node $N$, the union of all valid solutions permitted by its children is same as the set of valid solutions permitted at node $N$. Any invalid solution may be discarded while branching to ensure the algorithm terminates. This is satisfied while resolving precedence conflicts as discussed in Section 4.2. While resolving collision conflicts \cite{cbs}, the only solution invalidated is the one where agents are in collision due to the chosen conflicts. \hfill $\Box$

\noindent \textbf{[Theorem 1]} \textit{PC-CBS returns the optimal solution.}

\noindent \textit{Proof:} Since the high-level search explores solution costs in a best-first manner, the first goal node that is arrived in the CT will be the optimal solution. This results as a consequence of Lemma 1, which shows that $N.cost$ is a valid lower-bound for the best-first search and Lemma 2 which ensures completeness. \hfill $\Box$
\section{Hierarchical-CBS} \label{sec:hcbs}
\begin{figure}[t!]
    \centering
    \includegraphics[width=\linewidth]{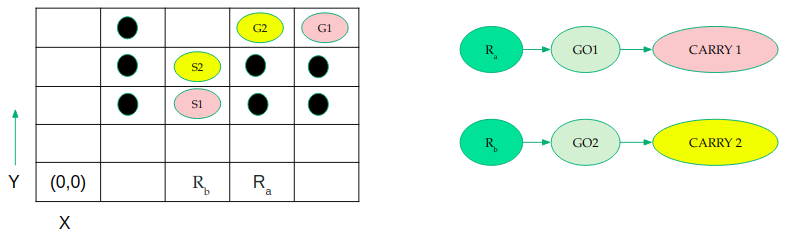}
    \vspace{-0.7cm}
    \caption{\small{An example PC-MAPF problem (left) and its Task Graph (right) where H-CBS is sub-optimal as $R_b$ needs to pre-emptively delay its path to achieve the optimal makespan cost solution (7) .}}
    \label{fig:sub-opt}
    \vspace{-0.5cm}
\end{figure}
\begin{figure*}[ht!]
\centering
\hspace*{\fill}
\begin{subfigure}{0.2\textwidth}
\centering
\includegraphics[width=\textwidth]{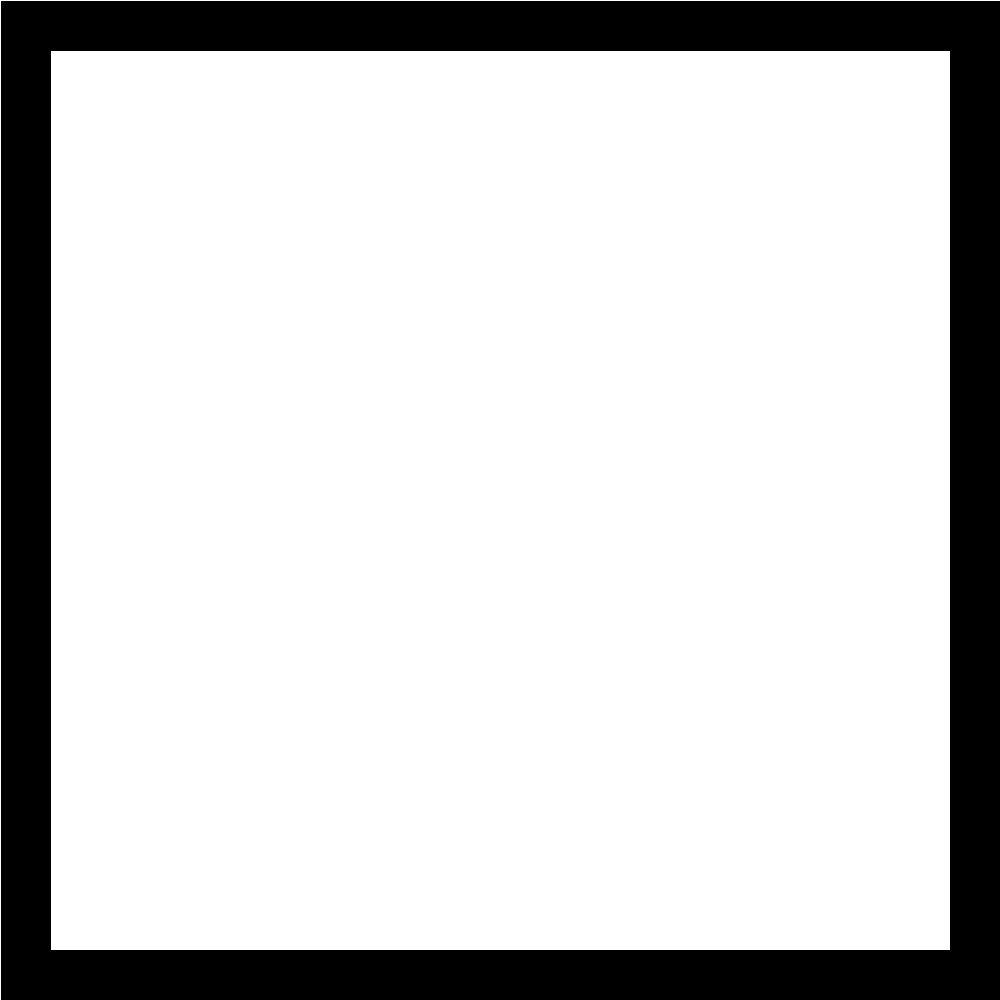}
\subcaption[]{\texttt{Empty Grid}}
\label{fig:env-pit}
\end{subfigure}
\hspace*{\fill}
\begin{subfigure}{0.2\textwidth}
\centering
\includegraphics[width=\textwidth]{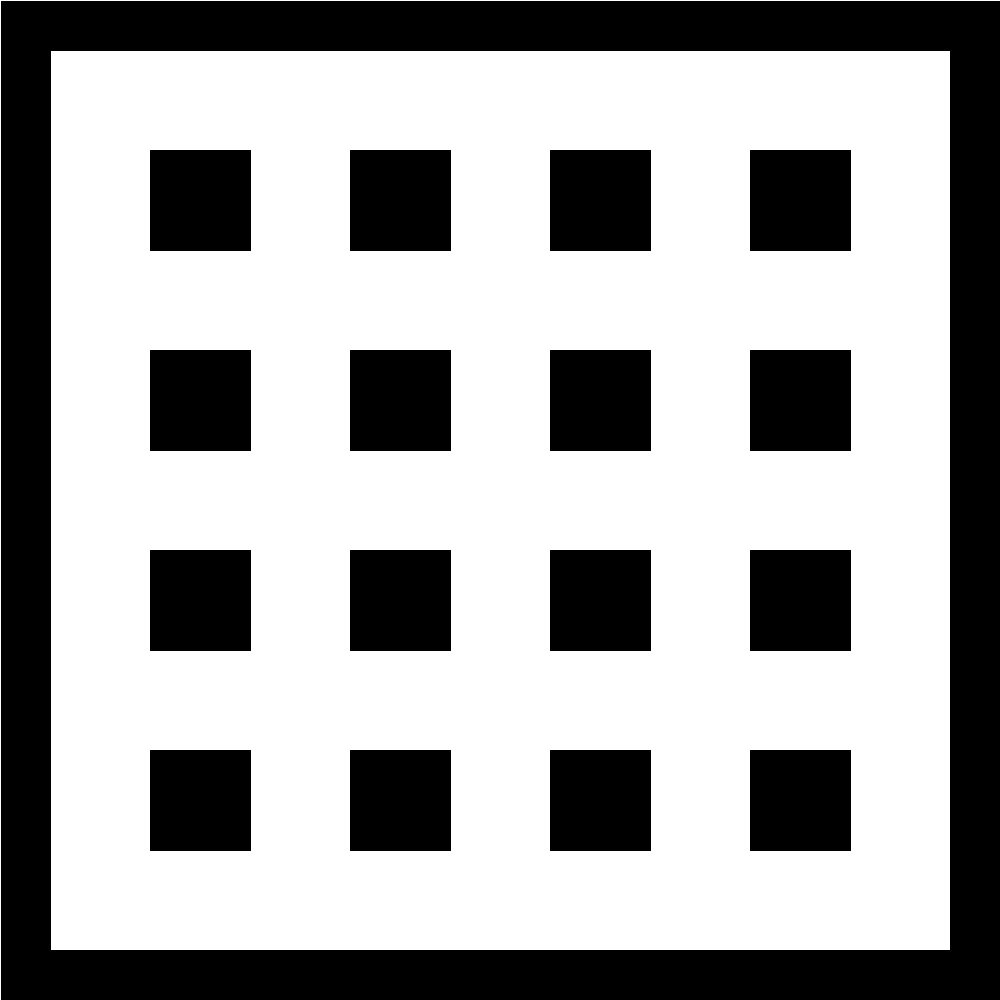}
\subcaption[]{\texttt{Warehouse Grid}}
\label{fig:env-hills}
\end{subfigure}
\hspace*{\fill}
\begin{subfigure}{0.2\textwidth}
\centering
\includegraphics[width=\textwidth]{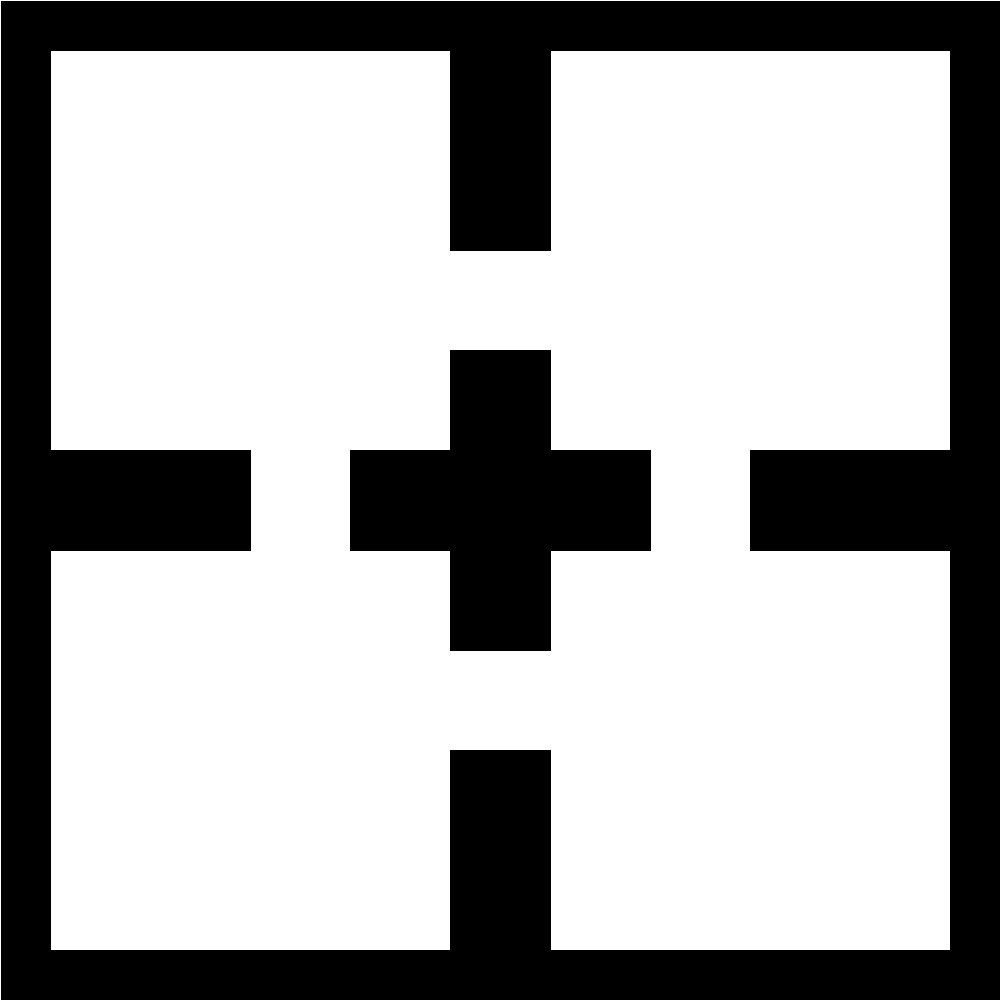}
\subcaption[]{\texttt{Maze-Gap}}
\label{fig:env-2hills}
\end{subfigure}
\hspace*{\fill}
\begin{subfigure}{0.2\textwidth}
\centering
\includegraphics[width=\textwidth]{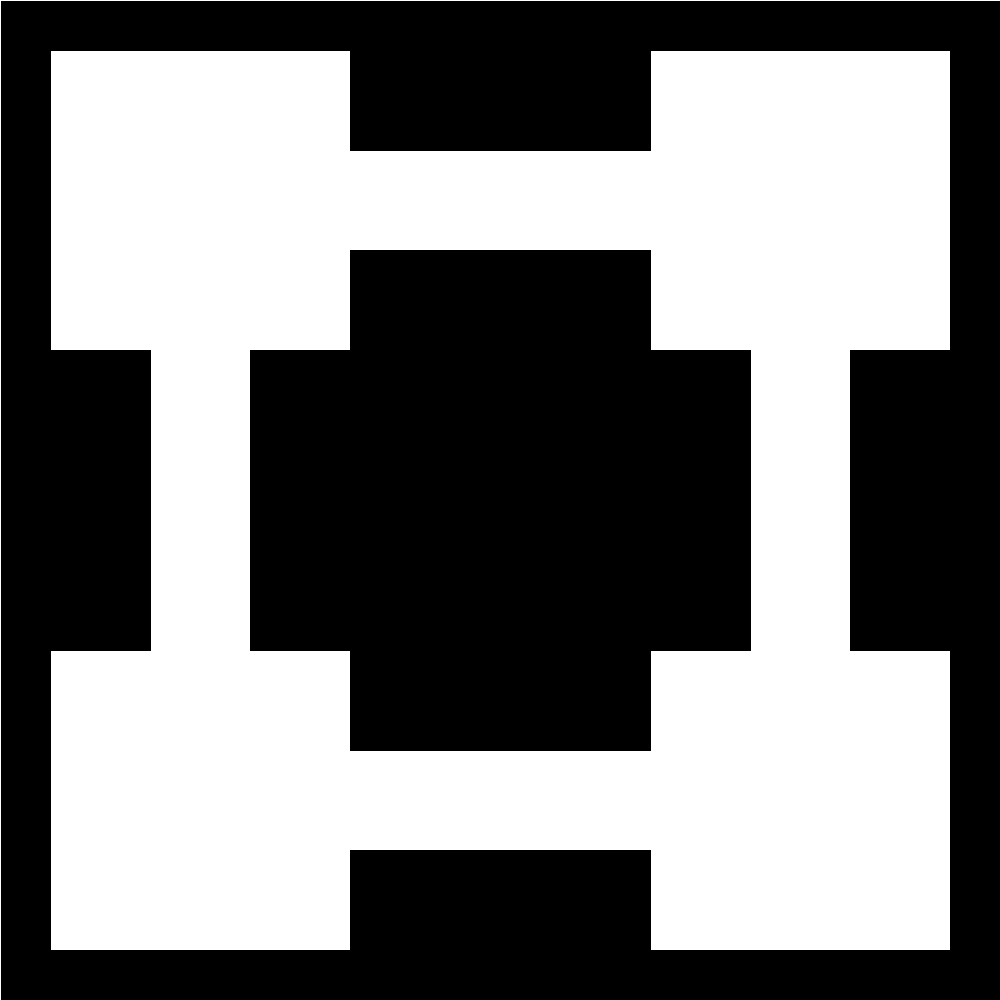}
\subcaption[]{\texttt{Maze-Tunnel}}
\label{fig:env-multihills}
\end{subfigure}
\hspace*{\fill}
\vspace{-0.25cm}
\caption{Visualization of 9x9 Grid Environments used for Warehouse Assembly and Collaborative MAPD Problems}
\label{fig:environments}
\vspace{-0.25cm}
\end{figure*}
\begin{table*}[ht!]
\centering
\caption{\label{assembly}Performance Metrics in Warehouse Assembly Problems over 4 Different Grid Environments}
\label{assembly}
\vspace{-0.25cm}
\begin{adjustbox}{width=2\columnwidth,center}
\begin{tabular}{|c|c|c|ccc|c|ccc|c|ccc|c|ccc|}
\hline
\multirow{2}{*}{\begin{tabular}[c]{@{}c@{}}Me-\\an\\Ta-\\sks\end{tabular}} & \multirow{2}{*}{\begin{tabular}[c]{@{}c@{}}No.\\of\\Age-\\nts\end{tabular}} & \multicolumn{4}{c|}{Empty Grid}                                                                                                                                                                                                    & \multicolumn{4}{c|}{Warehouse Grid}                                                                                                                                                                                                                                 & \multicolumn{4}{c|}{Maze-Gap}                                                                                                                                                                                                                                                        & \multicolumn{4}{c|}{Maze-Tunnel}                                                                                                                                                                                                                      \\ 
\cline{3-18}
                                                                           &                                                                             & \begin{tabular}[c]{@{}c@{}}PC-\\CBS\\Sol.\end{tabular} & \begin{tabular}[c]{@{}c@{}}H-\\CBS\\Sol.\end{tabular} & \begin{tabular}[c]{@{}c@{}}\% \\sub-\\opt.\end{tabular} & \begin{tabular}[c]{@{}c@{}}avg.\\reg-\\ret\end{tabular} & \begin{tabular}[c]{@{}c@{}}PC-\\CBS\\\textasciitilde{}Sol.\end{tabular} & \begin{tabular}[c]{@{}c@{}}H-\\CBS\\Sol.\end{tabular} & \begin{tabular}[c]{@{}c@{}}\% \\sub-\\opt\end{tabular} & \begin{tabular}[c]{@{}c@{}}avg.\\\textasciitilde{}reg-\\ret\end{tabular} & \begin{tabular}[c]{@{}c@{}}PC-\\CBS\\\textasciitilde{}Sol.\end{tabular} & \begin{tabular}[c]{@{}c@{}}H-\\CBS\\\textasciitilde{}Sol.\end{tabular} & \begin{tabular}[c]{@{}c@{}}\% \\sub-\\opt\end{tabular} & \begin{tabular}[c]{@{}c@{}}avg.\\\textasciitilde{}reg-\\ret\end{tabular} & \begin{tabular}[c]{@{}c@{}}PC-\\CBS\\\textasciitilde{}Sol.\end{tabular} & \begin{tabular}[c]{@{}c@{}}H-\\CBS\\Sol.\end{tabular} & \begin{tabular}[c]{@{}c@{}}\% \\sub-\\opt\end{tabular} & \begin{tabular}[c]{@{}c@{}}avg. \\reg-\\ret\end{tabular}  \\ 
\hline
\multirow{3}{*}{2}                                                         & 2                                                                           & 1.00                                                   & 1.00                                                  & 0\%                                                     & 0                                                       & 1.00                                                                    & 1.00                                                  & 0\%                                                    & 0                                                                        & 1.00                                                                    & 1.00                                                                   & 2\%                                                    & 0.02                                                                     & 0.99                                                                    & \textbf{1.00}                                         & 7\%                                                    & 0.14                                                      \\
                                                                           & 4                                                                           & 1.00                                                   & 1.00                                                  & 0\%                                                     & 0                                                       & 1.00                                                                    & 1.00                                                  & 4\%                                                    & 0.07                                                                     & 1.00                                                                    & 1.00                                                                   & 15\%                                                   & 0.23                                                                     & \textbf{0.96}                                                           & 0.93                                                  & 31\%                                                   & 0.61                                                      \\
                                                                           & 6                                                                           & 1.00                                                   & 1.00                                                  & 5\%                                                     & 0.08                                                    & 0.98                                                                    & \textbf{0.99}                                         & 9\%                                                    & 0.13                                                                     & \textbf{0.97}                                                           & 0.96                                                                   & 33\%                                                   & 0.59                                                                     & \textbf{0.89}                                                           & 0.44                                                  & 41\%                                                   & 0.68                                                      \\ 
\hline
\multirow{3}{*}{3}                                                         & 2                                                                           & 1.00                                                   & 1.00                                                  & 0\%                                                     & 0                                                       & 1.00                                                                    & 1.00                                                  & 1\%                                                    & 0.02                                                                     & 0.99                                                                    & \textbf{1.00}                                                          & 3\%                                                    & 0.05                                                                     & 1.00                                                                    & 1.00                                                  & 16\%                                                   & 0.29                                                      \\
                                                                           & 4                                                                           & 1.00                                                   & 1.00                                                  & 0\%                                                     & 0                                                       & \textbf{0.99}                                                           & 0.98                                                  & 9\%                                                    & 0.16                                                                     & \textbf{0.99}                                                           & 0.97                                                                   & 27\%                                                   & 0.45                                                                     & \textbf{0.89}                                                           & 0.68                                                  & 41\%                                                   & 0.75                                                      \\
                                                                           & 6                                                                           & 0.99                                                   & \textbf{1.00}                                                  & 3\%                                                     & 0.03                                                    & \textbf{0.96}                                                                    & 0.94                                                  & 11\%                                                   & 0.16                                                                     & 0.91                                                                    & 0.88                                                                   & 20\%                                                   & 0.34                                                                     & \textbf{0.62}                                                                    & 0.29                                                  & 39\%                                                   & 0.89                                                      \\ 
\hline
\multirow{3}{*}{4}                                                         & 2                                                                           & 1.00                                                   & 1.00                                                  & 2\%                                                     & 0.02                                                    & 1.00                                                                    & 1.00                                                  & 3\%                                                    & 0.04                                                                     & 1.00                                                                    & 1.00                                                                   & 9\%                                                    & 0.17                                                                     & 0.99                                                                    & 0.99                                                  & 21\%                                                   & 0.36                                                      \\
                                                                           & 4                                                                           & 1.00                                                   & 1.00                                                  & 2\%                                                     & 0.02                                                    & \textbf{1.00}                                                           & 0.96                                                  & 15\%                                                   & 0.22                                                                     & \textbf{0.99}                                                           & 0.98                                                                   & 21\%                                                   & 0.37                                                                     & \textbf{0.85}                                                           & 0.65                                                  & 39\%                                                   & 0.88                                                      \\
                                                                           & 6                                                                           & 0.99                                                   & \textbf{1.00}                                         & 3\%                                                     & 0.05                                                    & \textbf{0.89}                                                           & 0.85                                                  & 17\%                                                   & 0.19                                                                     & 0.71                                                                    & \textbf{0.76}                                                          & 20\%                                                   & 0.38                                                                     & \textbf{0.40}                                                           & 0.29                                                  & 50\%                                                   & 1.10                                                      \\
\hline
\multicolumn{1}{l}{}                                                       & \multicolumn{1}{l}{}                                                        & \multicolumn{1}{l}{}                                   & \multicolumn{1}{l}{}                                  & \multicolumn{1}{l}{}                                    & \multicolumn{1}{l}{}                                    & \multicolumn{1}{l}{}                                                    & \multicolumn{1}{l}{}                                  & \multicolumn{1}{l}{}                                   & \multicolumn{1}{l}{}                                                     & \multicolumn{1}{l}{}                                                    & \multicolumn{1}{l}{}                                                   & \multicolumn{1}{l}{}                                   & \multicolumn{1}{l}{}                                                     & \multicolumn{1}{l}{}                                                    & \multicolumn{1}{l}{}                                  & \multicolumn{1}{l}{}                                   & \multicolumn{1}{l}{}                                      \\
\multicolumn{1}{l}{}                                                       & \multicolumn{1}{l}{}                                                        & \multicolumn{1}{l}{}                                   & \multicolumn{1}{l}{}                                  & \multicolumn{1}{l}{}                                    & \multicolumn{1}{l}{}                                    & \multicolumn{1}{l}{}                                                    & \multicolumn{1}{l}{}                                  & \multicolumn{1}{l}{}                                   & \multicolumn{1}{l}{}                                                     & \multicolumn{1}{l}{}                                                    & \multicolumn{1}{l}{}                                                   & \multicolumn{1}{l}{}                                   & \multicolumn{1}{l}{}                                                     & \multicolumn{1}{l}{}                                                    & \multicolumn{1}{l}{}                                  & \multicolumn{1}{l}{}                                   & \multicolumn{1}{l}{}                                      \\
\multicolumn{1}{l}{}                                                       & \multicolumn{1}{l}{}                                                        & \multicolumn{1}{l}{}                                   & \multicolumn{1}{l}{}                                  & \multicolumn{1}{l}{}                                    & \multicolumn{1}{l}{}                                    & \multicolumn{1}{l}{}                                                    & \multicolumn{1}{l}{}                                  & \multicolumn{1}{l}{}                                   & \multicolumn{1}{l}{}                                                     & \multicolumn{1}{l}{}                                                    & \multicolumn{1}{l}{}                                                   & \multicolumn{1}{l}{}                                   & \multicolumn{1}{l}{}                                                     & \multicolumn{1}{l}{}                                                    & \multicolumn{1}{l}{}                                  & \multicolumn{1}{l}{}                                   & \multicolumn{1}{l}{}                                      \\
\multicolumn{1}{l}{}                                                       & \multicolumn{1}{l}{}                                                        & \multicolumn{1}{l}{}                                   & \multicolumn{1}{l}{}                                  & \multicolumn{1}{l}{}                                    & \multicolumn{1}{l}{}                                    & \multicolumn{1}{l}{}                                                    & \multicolumn{1}{l}{}                                  & \multicolumn{1}{l}{}                                   & \multicolumn{1}{l}{}                                                     & \multicolumn{1}{l}{}                                                    & \multicolumn{1}{l}{}                                                   & \multicolumn{1}{l}{}                                   & \multicolumn{1}{l}{}                                                     & \multicolumn{1}{l}{}                                                    & \multicolumn{1}{l}{}                                  & \multicolumn{1}{l}{}                                   & \multicolumn{1}{l}{}                                      \\
\multicolumn{1}{l}{}                                                       & \multicolumn{1}{l}{}                                                        & \multicolumn{1}{l}{}                                   & \multicolumn{1}{l}{}                                  & \multicolumn{1}{l}{}                                    & \multicolumn{1}{l}{}                                    & \multicolumn{1}{l}{}                                                    & \multicolumn{1}{l}{}                                  & \multicolumn{1}{l}{}                                   & \multicolumn{1}{l}{}                                                     & \multicolumn{1}{l}{}                                                    & \multicolumn{1}{l}{}                                                   & \multicolumn{1}{l}{}                                   & \multicolumn{1}{l}{}                                                     & \multicolumn{1}{l}{}                                                    & \multicolumn{1}{l}{}                                  & \multicolumn{1}{l}{}                                   & \multicolumn{1}{l}{}                                      \\
\multicolumn{1}{l}{}                                                       & \multicolumn{1}{l}{}                                                        & \multicolumn{1}{l}{}                                   & \multicolumn{1}{l}{}                                  & \multicolumn{1}{l}{}                                    & \multicolumn{1}{l}{}                                    & \multicolumn{1}{l}{}                                                    & \multicolumn{1}{l}{}                                  & \multicolumn{1}{l}{}                                   & \multicolumn{1}{l}{}                                                     & \multicolumn{1}{l}{}                                                    & \multicolumn{1}{l}{}                                                   & \multicolumn{1}{l}{}                                   & \multicolumn{1}{l}{}                                                     & \multicolumn{1}{l}{}                                                    & \multicolumn{1}{l}{}                                  & \multicolumn{1}{l}{}                                   & \multicolumn{1}{l}{}                                      \\
\multicolumn{1}{l}{}                                                       & \multicolumn{1}{l}{}                                                        & \multicolumn{1}{l}{}                                   & \multicolumn{1}{l}{}                                  & \multicolumn{1}{l}{}                                    & \multicolumn{1}{l}{}                                    & \multicolumn{1}{l}{}                                                    & \multicolumn{1}{l}{}                                  & \multicolumn{1}{l}{}                                   & \multicolumn{1}{l}{}                                                     & \multicolumn{1}{l}{}                                                    & \multicolumn{1}{l}{}                                                   & \multicolumn{1}{l}{}                                   & \multicolumn{1}{l}{}                                                     & \multicolumn{1}{l}{}                                                    & \multicolumn{1}{l}{}                                  & \multicolumn{1}{l}{}                                   & \multicolumn{1}{l}{}                                     
\end{tabular}
\end{adjustbox}
\vspace{-2.8cm}
\end{table*}
\begin{table*}[ht!]
\centering
\caption{\label{cmapd}Performance Metrics in Collaborative MAPD Problems over 4 Different Grid Environments}
\label{cmapd}
\vspace{-0.25cm}
\begin{adjustbox}{width=2\columnwidth,center}
\begin{tabular}{|c|c|c|c|ccc|c|ccc|c|ccc|c|ccc|}
\hline
\multirow{2}{*}{\begin{tabular}[c]{@{}c@{}}Col.\\Deg\\-ree\end{tabular}} & \multirow{2}{*}{\begin{tabular}[c]{@{}c@{}}Me-\\an\\Ta-\\sks\end{tabular}} & \multirow{2}{*}{\begin{tabular}[c]{@{}c@{}}No.\\of\\age-\\nts\end{tabular}} & \multicolumn{4}{c|}{Empty Grid}                                                                                                                                                                                                    & \multicolumn{4}{c|}{Warehouse Grid}                                                                                                                                                                                                  & \multicolumn{4}{c|}{Maze-Gap}                                                                                                                                                                                                        & \multicolumn{4}{c|}{Maze-Tunnel}                                                                                                                                                                                                        \\ 
\cline{4-19}
                                                                         &                                                                            &                                                                             & \begin{tabular}[c]{@{}c@{}}PC-\\CBS\\Sol.\end{tabular} & \begin{tabular}[c]{@{}c@{}}H-\\CBS\\Sol.\end{tabular} & \begin{tabular}[c]{@{}c@{}}\% \\sub-\\opt.\end{tabular} & \begin{tabular}[c]{@{}c@{}}avg.\\reg-\\ret\end{tabular} & \begin{tabular}[c]{@{}c@{}}PC-\\CBS \\Sol.\end{tabular} & \begin{tabular}[c]{@{}c@{}}H-\\CBS \\Sol.\end{tabular} & \begin{tabular}[c]{@{}c@{}}\% \\sub-\\opt\end{tabular} & \begin{tabular}[c]{@{}c@{}}avg. \\reg-\\ret\end{tabular} & \begin{tabular}[c]{@{}c@{}}PC-\\CBS \\Sol.\end{tabular} & \begin{tabular}[c]{@{}c@{}}H-\\CBS \\Sol.\end{tabular} & \begin{tabular}[c]{@{}c@{}}\% \\sub-\\opt\end{tabular} & \begin{tabular}[c]{@{}c@{}}avg. \\reg-\\ret\end{tabular} & \begin{tabular}[c]{@{}c@{}}PC-\\CBS \\Sol.\end{tabular} & \begin{tabular}[c]{@{}c@{}}H-\\CBS \\Sol.\end{tabular} & \begin{tabular}[c]{@{}c@{}}\% \\sub-\\opt.\end{tabular} & \begin{tabular}[c]{@{}c@{}}avg. \\reg-\\ret\end{tabular}  \\ 
\hline
\multirow{4}{*}{2}                                                       & \multirow{2}{*}{2}                                                         & 4                                                                           & 1.00                                                   & 1.00                                                  & 2\%                                                     & 0.02                                                    & 1.00                                                    & 1.00                                                   & 4\%                                                    & 0.04                                                     & 1.00                                                    & 1.00                                                   & 4\%                                                    & 0.05                                                     & 0.99                                                    & 0.99                                                   & 16\%                                                    & 0.29                                                      \\
                                                                         &                                                                            & 6                                                                           & 1.00                                                   & 1.00                                                  & 6\%                                                     & 0.06                                                    & 0.83                                                    & \textbf{0.98}                                          & 7\%                                                    & 0.07                                                     & 0.96                                                    & \textbf{1.00}                                          & 14\%                                                   & 0.20                                                     & 0.84                                                    & \textbf{0.92}                                          & 30\%                                                    & 0.44                                                      \\ 
\cline{2-19}
                                                                         & \multirow{2}{*}{3}                                                         & 4                                                                           & 1.00                                                   & 1.00                                                  & 6\%                                                     & 0.07                                                    & 0.73                                                    & \textbf{1.00}                                          & 11\%                                                   & 0.12                                                     & 0.93                                                    & \textbf{1.00}                                          & 14\%                                                   & 0.20                                                     & 0.78                                                    & \textbf{0.98 }                                         & 12\%                                                    & 0.31                                                      \\
                                                                         &                                                                            & 6                                                                           & 0.99                                                   & \textbf{1.00 }                                        & 8\%                                                     & 0.08                                                    & 0.42                                                    & \textbf{0.99 }                                         & 24\%                                                   & 0.29                                                     & 0.66                                                    & 1.00                                                   & 28\%                                                   & 0.37                                                     & 0.54                                                    & \textbf{0.90}                                          & 45\%                                                    & 0.68                                                      \\ 
\hline
\multirow{4}{*}{3}                                                       & \multirow{2}{*}{2}                                                         & 3                                                                           & 1.00                                                   & 1.00                                                  & 0\%                                                     & 0.00                                                    & 1.00                                                    & 1.00                                                   & 0\%                                                    & 0.00                                                     & 1.00                                                    & 1.00                                                   & 0\%                                                    & 0.00                                                     & 1.00                                                    & 1.00                                                   & 0\%                                                     & 0.00                                                      \\
                                                                         &                                                                            & 6                                                                           & 0.93                                                   & \textbf{0.99 }                                        & 2\%                                                     & 0.02                                                    & 0.72                                                    & \textbf{1.00 }                                         & 1\%                                                    & 0.01                                                     & 0.73                                                    & \textbf{0.96 }                                         & 15\%                                                   & 0.17                                                     & 0.65                                                    & \textbf{0.90 }                                         & 12\%                                                    & 0.13                                                      \\ 
\cline{2-19}
                                                                         & \multirow{2}{*}{3}                                                         & 3                                                                           & 0.15                                                   & \textbf{0.96 }                                        & 0\%                                                     & 0.00                                                    & 0.20                                                    & \textbf{1.00 }                                         & 0\%                                                    & 0.00                                                     & 0.11                                                    & \textbf{0.88 }                                         & 0\%                                                    & 0.00                                                     & 0.26                                                    & \textbf{1.00 }                                         & 0\%                                                     & 0.00                                                      \\
                                                                         &                                                                            & 6                                                                           & 0.20                                                   & \textbf{0.98 }                                        & 30\%                                                    & 0.40                                                    & 0.13                                                    & \textbf{0.92 }                                         & 33\%                                                   & 0.50                                                     & 0.13                                                    & \textbf{0.91 }                                         & 23\%                                                   & 0.31                                                     & 0.12                                                    & \textbf{0.77 }                                         & 36\%                                                    & 0.45                                                  \\
\hline
\end{tabular}
\end{adjustbox}
\end{table*}

In this section, we briefly explain Hierarchical-CBS (H-CBS) \cite{hcbs}, a recently proposed algorithm that solves the PC-MAPF problem and demonstrate an example case where H-CBS fails to find an optimal path. In its pre-processing step, H-CBS finds the time required for completion of each of the tasks without taking into account collisions between agents. It then calculates the \textit{slack} available at each task by calculating the amount of extra time that can be spent in doing that task without increasing the global makespan cost of the solution. To find a valid solution to the PC-MAPF problem, H-CBS performs a three-level hierarchical search which involve in order of level from top to down: CBS, Incremental Slack Prioritized Search (ISPS) and A*. At the top level, CBS is used to detect collision conflicts between plans of individual tasks and resolve them by propagating collision constraints to the lower level algorithms. Given a set of constraints, ISPS traverses the precedence constrained task graph in topological order, and calls A* to find the shortest collision-free plan of each task segment prioritized by the amount of slack available to a task. 

Consider the PC-MAPF problem presented in Fig \ref{fig:sub-opt}. This denotes a pickup and delivery problem where, agent $R_a$ must go to the position S1 (denoted by a GO-1 task), and then pickup, carry object 1 over and deliver it at G1 (denoted by a CARRY-1 task), and agent $R_b$ must go to the position S2 (denoted by a GO-2 task), and then pickup, carry object 2 over and deliver it at G2 (denoted by a CARRY-2 task). The Task Graph shown in the figure consists of the four tasks: GO-1, CARRY-1, GO-2 and CARRY-2 and the in-schedule precedence constraints between these tasks are denoted by directed edges. The optimal solution requires $R_b$ to pre-emptively delay its path, allowing $R_a$ to enter the narrow passage first and has a makespan cost of 7.  In the first iteration of H-CBS, paths are computed without considering collisions. When these paths are collision checked, a collision conflict is found at timestep 6 at G2. However, this conflict is between tasks CARRY-1 and CARRY-2. In fact, conflicts always arise between these two tasks in future iterations. The GO tasks have no collisions and do not need to replan their paths. Hence, in the eventual solution found by H-CBS, the agents will greedily enter ($R_b$ first) the passage and then have to come out again to reorder themselves in the correct order ($R_a$ first). Since the makespan of this solution is 11, it is an example of a class of problems in which H-CBS cannot guarantee optimality.
\vspace{-3.2mm}
\section{Experimental Results} \label{sec:experiment}

The PC-CBS and H-CBS algorithms are implemented in C++ using the Boost Graph library \cite{boost} for standard graph operations. Experiments are run on an Intel Core i7-1165G7 CPU @ 2.80GHz×8 processor with 15.3 GB of memory. The timeout is set to 300s for each run on an individual problem instance. Results are shown over four 9x9 grid-worlds (Fig. \ref{fig:environments}). We consider the following performance metrics in our tables. The last two metrics are computed over problems solved by both algorithms. 
\begin{enumerate}
\vspace{-0.7mm}
    \item \textbf{ (PC-CBS, H-CBS)\_Sol.} is defined as the number of problems solved by an algorithm divided by the total number of problems. (100 for each problem set)
    \item \textbf{\% sub-opt. } is defined as the number of problems where H-CBS is sub-optimal divided by the number of problems solved by both algorithms. (expressed as a percentage)
    \item \textbf{Avg. regret} is the mean surplus makespan, the amount of extra timesteps needed by H-CBS compared to the optimal makespan found by PC-CBS. ( $= 1/(Solved\_$ $by\_both) \sum H\-CBS_{mkspn}-PC\-CBS_{mkspn}$)
\vspace{-1.0mm}
\end{enumerate}

Cells which are colored black represent obstacles and cannot be occupied by any agents. At any timestep, an agent may either \textit{wait} at its current cell, or \textit{move} to its four-connected neighbourhood in the grid. The environments represent diverse scenarios to evaluate the optimality of our algorithm. The \texttt{empty-grid} does not contain any obstacles, providing maximum free space to agents to move in while the \textit{maze-tunnel} is highly constricted requiring passage through narrow tunnels to move between quadrants. We choose two domains of PC-MAPF problems for testing our algorithm. \textbf{Warehouse Assembly} problems are a special class of PC-MAPF where only one agent is assigned per task. \textbf{Collaborative MAPD} problems are a special class of PC-MAPF where precedence constraints arise only bcecause of multiple agents being assigned to pickup and delivery tasks. Each problem set consists of 100 randomly generated instances. The results in Table \ref{assembly} and Table \ref{cmapd} show these performance metrics over different number of agents in the grid and mean number of tasks alloted to each agent. Additionally for collaborative MAPD, the number of agents required for each task (collaboration degree) is varied between 2 and 3.

We assign random initial locations for each agent, and random pickup \& delivery locations for transport tasks to create problems. The Task Graph edges, representing precedence constraints between tasks, are randomly generated for warehouse assembly problems. Agents are allotted to their task sequences using a greedy assignment scheme. Among each task that can begin, we consider the set of agents that can reach it's start location. The task which can be started earliest is picked for allocation. This assignment strategy does not consider collisions between agents but it respects the precedence constraints between tasks.

From the results in Table \ref{assembly}, it is observed that PC-CBS not only finds optimal makespan paths but also matches or outperforms (marked in bold) H-CBS in \textbf{30/36} of the problem sets chosen for evaluation. In spite of PC-CBS computing longer A* searches, it is seen that the sub-optimality of H-CBS makes it prone to getting stuck in large search trees, especially in the \texttt{maze} environments which are similar to the example shown in Fig. \ref{fig:sub-opt}. The surplus makespan (regret) and the percentage of problems where H-CBS is sub-optimal increases with the complexity of the environment. In the \texttt{empty-grid}, H-CBS is almost always close to the optimal cost which is expected since planning individual tasks in a free region will avoid situations requiring pre-emptive delays. Despite the \texttt{warehouse-grid} having more obstacle cells than the \texttt{maze-gap} environment, it is generally seen that solutions found by H-CBS are closer to optimal in this environment. The existence of multiple equal quality pathways from one cell to another helps the segmented planner, whereas in the \texttt{maze-gap} environment, the agents are constricted to move through gaps to move across quadrants. The bottleneck problem is only exaggerated in the \texttt{maze-tunnel} environment where the percentage of sub-optimal solutions goes as high as $50\%$.

\begin{figure}[t!]
    \centering
    \includegraphics[width=\linewidth]{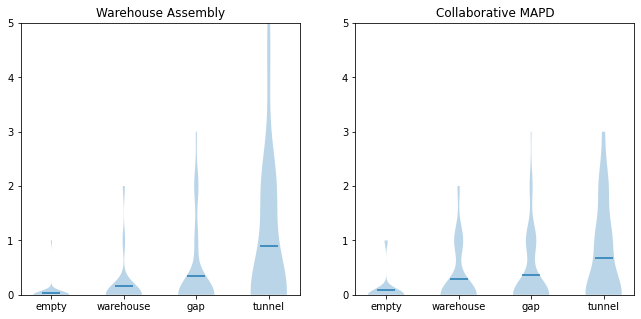}
    \vspace{-0.7cm}
    \caption{\small {Surplus makespan, i.e, regret of H-CBS is visualized over the different gridworld environments using a Violin plot. The thick line in the middle indicates the mean regret. The problem set chosen here consisted of 6 agents with 3 mean tasks for both problem domains. The degree of collaboration is 2 for Collaborative MAPD.}}
    \label{fig:regret}
\vspace{-0.5cm}
\end{figure}

Table \ref{cmapd} highlights the benefit of segmented planning in finding feasible solution paths for collaborative MAPD tasks. H-CBS matches or outperforms PC-CBS in finding more feasible solutions in all problem-sets. The amount of sub-optimality, however, is still significantly large in some case and shows the same variation with environments as shown in Fig \ref{fig:regret}. The problem sets where three agents collaborate for pickup tasks requiring three agents show no difference in makespan in any of the problems. This is expected since pre-emptive delays are never required in the case. Collaborative transport can start only when all agents arrive at the pickup location together. It is also observed that the performance of PC-CBS degrades quickly with increasing number of agents assigned to a task. For collaborative tasks, a cluster of conflicts can form around the pickup and delivery. Since the low-level search of PC-CBS employs a longer multi-waypoint search, the number of conflicts grows exponentially faster than in H-CBS.

\vspace{-0.25cm}
\subsection*{Extensions}
While PC-CBS has been proposed for precedence constrained tasks, the algorithmic framework can be adapted for other types of cross-schedule dependencies like synchronization and time-windowing constraints. Since the valid duration for tasks are represented by intervals, PC-CBS can integrate Safe-Interval Path Planning (SIPP) \cite{sipp} in its low-level search and be extended to continuous time domains similar to the work by \cite{ccbs}.  Recent work in improving CBS have significantly improved its run-time by choosing promising conflicts \cite{icbs}. Other works \cite{subopt_cbs} have proposed a highly efficient and bounded sub-optimal version of the algorithm. \cite{disjoint_cbs} found an order of magnitude improvement by incorporating positive constraints, which act as waypoints, to resolve collision conflicts in CBS. All of these improvements can be applied to PC-CBS, especially in the context of precedence conflicts.

\section{Conclusion} \label{sec:conclusion}

This work proposed an optimal algorithm for PC-MAPF based on a state-of-the-art MAPF solver, CBS. We represent warehouse assembly and collaborative MAPD problems as a Task Graph, which is input to our solver. We prove the optimality of our algorithm and evaluate the sub-optimality of another recently proposed PC-MAPF solver. Next steps include integration of a task assignment module on top of the path planner. The use of a multi-goal variant of A* can be used to find paths for a set of unordered tasks assigned to each agent reducing the burden on the Task Allocation search problem. Another promising direction of work includes the combination of collaborative transport tasks with precedence constraints. In the future, we also intend to explore design of intelligent warehouses through PC-CBS, which is pertinent to recent MAPF research \cite{bluesky}.


\bibliographystyle{IEEEtran}
\bibliography{sample}


\end{document}